\documentclass{article}
\begin{document}
\title{Matter waves in a gravitational field: \\
An index of refraction for massive particles \\
in general relativity}

\normalsize
\author{James Evans \\ 
Department of Physics, University of Puget Sound, \\
Tacoma, WA 98416 \\
jcevans@ups.edu
\and 
Paul~M.~Alsing \\ Albuquerque High Performance Computing Center,\\
University of New Mexico, Albuquerque, NM 87131 \\
alsing@ahpcc.unm.edu
\and
Stefano Giorgetti \\
Department of Physics, University of Milan, \\
20133 Milano, Italy\\
epo@telemacus.it
\and
Kamal Kanti Nandi \\Department of Mathematics, University of North Bengal,\\
Darjeeling (WB) 734430, India\\
kamalnandi@hotmail.com}

\maketitle

\begin{abstract}
We consider the propagation of massive-particle de Broglie waves in a static, isotropic metric in general relativity.  We demonstrate the existence of an index of refraction that governs the waves and that has all the properties of a classical index of refraction. We confirm our interpretation with a WKB solution of the general-relativistic Klein-Gordon equation.  Finally, we make some observations on the significance of the optical action.
\end{abstract}

\newpage
\section*{I. INTRODUCTION}

In the case of static, isotropic metrics, the optical-mechanical analogy leads to a remarkable simplification of the geodesic equation of motion.  The trajectories of both massive and massless particles can be parameterized by one and the same parameter, the optical action $A$.  Moreover, for both photons and massive particles, the equation of motion assumes the form of Newton's second law.  Thus the exact general-relativistic problem becomes formally identical to the classical-mechanical or, if one prefers, to the classical geometrical-optical, problem.

This formalism also provides an easy way to introduce quantum-mechanical wave packets into the context of general relativity. In this paper we consider the propagation of massive-particle de Broglie waves in a gravitational field represented by a static, isotropic metric.  We shall use simple, but rigorous, semi-classical arguments to demonstrate the existence of an index of refraction that governs the waves. Our expression for the index of refraction will be confirmed by a WKB solution of the general-relativistic Klein-Gordon equation.  We shall see that the same index of refraction may be used in both the classical-particle and wave-mechanical pictures of propagation.  Finally, this treatment will shed new light on the physical significance of the optical action.  

\subsection*{A. The gravitational field as an optical medium}
Our point of departure will be an exact formulation of the geodesic equation of motion which is nevertheless of Newtonian form.  The remainder of the Introduction will summarize this formulation of the geodesic problem.  This summary will serve not only to fix the notation but also to introduce readers to the optical action, the physical quantity that plays the central role in all that follows.  Throughout this paper, we suppose that the metric is static and isotropic, so that the line element is of the form 
\begin{equation}
\label{metric}
ds^{2} = \Omega^{2}({\bf r})c_{0}^{2}dt^{2} - \Phi^{-2}({\bf r})|d{\bf r}|^{2},
\end{equation}
where $\Omega$ and $\Phi$ are functions of the spatial coordinates ${\bf r} = 
(r, \theta, \phi)$ or $(x, y, z)$ and where $c_{0}$ is the vacuum speed of light.  Many metrics of physical interest can be put into this form, including the Schwarzschild metric.\cite{noteiso} The isotropic coordinate speed of light $c({\bf r})$ is determined by the condition that the geodesics be null ($ds = 0):$
\begin{equation}
c({\bf r}) = \left|\frac{d{\bf r}}{dt}\right| = c_{0}\Phi({\bf r})\Omega({\bf 
r}).
\end{equation}
Thus the effective index of refraction for light in the gravitational field is\cite{noteEddington} 
\begin{equation}
n({\bf r}) = \Phi^{-1}\Omega^{-1}.
\end{equation}

\subsection*{B. Relativistic geometrical optics in Newtonian form}

This index of refraction may be used in any formulation of geometrical optics.  For example, the shapes of light rays are governed by Fermat's principle:
\begin{equation}
\delta \int_{{\bf x}_{1}}^{{\bf x}_{2}} \, n \, dl = 0.
\end{equation}
$\delta$ represents a variation of the integral produced by varying the path between two fixed points in space, ${\bf x}_{1}$ and ${\bf x}_{2}$.  $dl = |d{\bf r}|$ is the element of the path of integration in the three-dimensional space. From Eq.~(4) one may derive other formulations of geometrical optics, including the ``$F = ma$'' formulation.  In this formulation, the equation of the ray assumes the form of Newton's law of motion: 
\begin{equation}
\frac{d^{2}{\bf r}}{dA^{2}} = \nabla \left( \frac{1}{2} n^{2}c_{0}^{2} \right). \phantom{XXXXXXXX} {\rm (light)}
\end{equation}
This is an exact formulation of geometrical optics that was originally applied to classical isotropic media.\cite{noteER} However it is equally valid for static, isotropic 
metrics in general relativity.\cite{Nan95} Eq.~(5) reduces relativistic geometrical optics to the form of classical Newtonian mechanics.  The left side of Eq.~(5) has the form of an acceleration.  The right side has the form of a force and the effective potential-energy function is $- n^{2}c_{0}^{2}/2$.  The key feature of this formulation of geometrical optics is the use of the optical action $A$ as independent variable.  $A$ is defined by 
\begin{equation}
\left|\frac{d{\bf r}}{dA}\right| = nc_{0}. \phantom{XXXXXXXX} {\rm (light)}
\end{equation}
$dA$ is related in a simple way to the coordinate time $dt$ elapsed as light moves an 
infinitesimal distance $dl$ along the ray: since $dA = dl/nc_{0}$ and 
$dl = c_{0}dt/n$, 
\begin{equation}
dA = dt/n^{2}.
\end{equation}

\subsection*{C. Relativistic particle mechanics in Newtonian form}
The orbits of massive particles are obtained by requiring that they be 
geodesics:\begin{equation}
\delta \int_{{\bf x}_{1}, t_{1}}^{{\bf x}_{2}, t_{2}} ds = 0.
\end{equation}
Here $\delta$ represents a variation of the integral produced by varying the path between two fixed points in spacetime, $({\bf x}_{1}, t_{1})$ and $({\bf x}_{2}, t_{2})$.  We wish to cast this variational principle into the form of Eq.~(4) in order to make the optical-mechanical analogy as transparent as possible.\cite{noteENI1, noteENI2, noteAlsing} With the use of Eq.~(1) and (3), Eq.~(8) can be written in a form analogous to Hamilton's principle:
\begin{equation}
\delta \int_{t_{1}}^{t_{2}} L(x_{i}, V_{i} )\, dt = 0,
\end{equation}
where the effective Lagrangian is 
\begin{equation}
L(x_{i}, V_{i} ) = -mc_{0}^{2} \Omega \left[1 - V^{2}n^{2}/c_{0}^{2}\right]^{1/2}.
\end{equation}
In the Lagrangian, $\Omega$ and $n$ are functions of the spatial coordinates alone, 
$V^{2} = \sum_{i = 1}^{3} V_{i}^{2}$ if we choose to work in Cartesian coordinates, and $V_{i}$ denotes $dx_{i}/dt$.  The factor of the rest mass $m$ has been included for dimensional convenience.

The canonical momenta are
\begin{equation}
p_{i} = m\Omega n^{2} \left[1 - V^{2}n^{2}/c_{0}^{2}\right]^{-
1/2}V_{i}.
\end{equation}
The effective Hamiltonian is 
\begin{equation}
\label{H1}
H = mc_{0}^{2} \Omega \left[1 - V^{2}n^{2}/c_{0}^{2}\right]^{-1/2},
\end{equation}
or, if expressed in terms of the momenta,
\begin{equation}
\label{H2}
H = mc_{0}^{2}\left[\Omega^{2} + p^{2}/n^{2}m^{2}c_{0}^{2}\right]^{1/2}.
\end{equation}
$H$ is a constant of the motion, whose value may be calculated from Eq.~(12) if the particle coordinate speed $V$ is known at one point on the path. A little algebra applied to Eq.~(12) gives the speed function $V({\mathbf r})$:
\begin{equation}
V({\bf r}) = c_{0}n^{-1}\left[1 - m^{2}c_{0}^{4}\Omega^{2}/H^{2} \right]^{1/2}.
\end{equation}
Eq.~(14), for massive particles, corresponds to the familiar $v = c_{0}n^{-1}$ for light.  

From Hamilton's principle, Eq.~(9), one may derive Jacobi's form of Maupertuis's principle:
\begin{equation}
\delta \int_{{\bf x}_{1}}^{{\bf x}_{2}} p \, dl = 0,
\end{equation}
where now the path of integration though three-dimensional space is varied, subject to conservation of energy, between two fixed points in space, ${\bf x}_{1}$ and ${\bf x}_{2}$, but the times at the end points need not be held fixed. From Eq.~(11) and Eq.~(12) we have
\begin{equation}
p_{i} = Hn^{2}V_{i}/c_{0}^{2}
\end{equation}
and thus Eq.~(15) can be written 
\begin{equation}
\delta \int_{{\bf x}_{1}}^{{\bf x}_{2}} \frac{Hn^{2}V}{c_{0}^{2}} \, dl = 0.
\end{equation}
Here $n^{2}V$ is to be regarded as a function of the spatial coordinates. 

Eq.~(17), which governs the orbits of massive particles in static, isotropic metrics, is formally analogous to Maupertuis's principle, which is a suitable foundation for the Newtonian mechanics of a point-particle in a static, isotropic potential.  But Eq.~(17) is also analogous to Fermat's principle, Eq.~(4), which governs the shape of light rays in isotropic media.  Thus, corresponding to any formulation of geometrical optics that is derivable from Eq.~(4), there is a formulation of relativistic particle mechanics derivable from Eq.~(17).  

Indeed, the exact general-relativistic equation of motion for a massive particle may be written 
\begin{equation}
\frac{d^{2}{\bf r}}{dA^{2}} = \nabla \left( \frac{1}{2} n^{4}V^{2}\right). \phantom{XX} {\rm (massive\phantom{x}particle)}
\end{equation}
This equation is precisely of the form of Newton's second law:  acceleration = negative gradient of potential energy.  Note that, because $V = c_{0}/n$ for light, Eq.~(18) reduces to Eq.~(5) as a special case. 

The independent variable $A$ playing the role of the time in Eq.~(18) is the optical action, defined by
\begin{equation}
\left|\frac{d{\bf r}}{dA}\right| = n^{2}V. \phantom{XX} {\rm (massive\phantom{x}particle)}
\end{equation}
But since $V = |d{\bf r}/dt|$ it follows that, along the trajectory,
\begin{equation}
dA = dt/n^{2}.
\end{equation}
Thus the same independent variable $A$ is used to parameterize the path, whether the particle be massive or massless.  The only difference in the treatment of massive and massless particles is in the form of the potential energy function, $U({\bf r})= -n^{2}c_{0}^{2}/2$ or 
$-n^{4}V^{2}/2$, which appears under the gradient operator in Eq.~(5) or Eq.~(18). 

\section*{II.  DE BROGLIE WAVES IN THE \\GRAVITATIONAL MEDIUM}

With these results in hand, we are ready to introduce the de Broglie waves of a massive particle into the medium that represents a static, gravitational field. Let these waves have phase speed $v_{p}({\bf r})$, wave number $k({\bf r})$ and angular frequency $\omega$, related by $v_{p} = \omega /k$. Because of the similarity of Eq.~(4) and Eq.~(17) (as well as the similarity of Eq.~(5) and Eq.(18), we might speculate that $N = n^{2}V/c_{0}$ plays the role of an index of refraction for the massive particles.  We shall see that this is indeed the case.   

In the geometrical-optics limit, it is possible to speak of a ray, which is the trajectory of the point-particle.  The ray will be the path satisfying Fermat's principle, which we write in the form:
\begin{equation}
\delta \int_{{\bf x}_{1}}^{{\bf x}_{2}} k \, dl = 0.
\end{equation}
And now we may reason in exactly the same way as Louis de Broglie in his thesis of 1924.\cite{noteBro} The paths predicted by Eq.~(17) and Eq.~(21) will be the same if the two integrands are the same functions of the spatial coordinates.  However, they may differ by a multiplicative factor $f$.  Thus we require
\begin{equation}
Hn^{2}V/c_{0}^{2} = f k.
\end{equation}

\subsection*{A. The de Broglie wavelength of a massive particle}

To determine $f$ let us pass to the empty-space limit in which $H \rightarrow 
mc_{0}^{2}\gamma = mc_{0}^{2} (1 - V^{2}/ c_{0}^{2})^{-1/2}$ and $n^{2} \rightarrow 1$.  Eq.~(22)  becomes $mV \gamma = fk$.  This will be precisely the de Broglie relation of a relativistic free particle if $f = \hbar = h/2\pi$, where $h$ is Planck's constant.  Thus we have 
\begin{equation}
\label{Hn2V}
Hn^{2}V/c_{0}^{2} = \hbar k,
\end{equation}
which is just what we would have obtained if we had simply put 
\begin{equation}
p = \hbar k
\end{equation}
using Eq.~(16).  

Since $(H/c_{0}, {\bf p})$ and $(\omega/c_{0}, {\bf k})$ must each transform as a four-vector, we must also have
\begin{equation}
\label{hbarw}
H = \hbar \omega.
\end{equation}
Thus, given a wave of angular frequency $\omega$, the frequency remains constant along the whole trajectory or ray.  For a particle in an unbound orbit, $\omega$ is everywhere equal to the value of $\omega$ at spatial infinity. The optics of de Broglie waves in a static, isotropic metric is therefore analogous to the wave optics of classical isotropic media.  Of course, it must be emphasized that $\omega$ is a {\em coordinate frequency}, and not a directly measurable frequency.  $\omega$ is $2\pi$ times the number of wave crests crossing through a fixed point ${\mathbf r}$ per unit coordinate time $t$.  From Eq.~(\ref{metric}), the frequency that would be measured by an observer at ${\bf r}$ is $\tilde{\omega} = \omega/ \Omega({\bf r})$.

From Eq.~(\ref{Hn2V}), the wavelength of the massive particle in the gravitational field is
\begin{equation}
\lambda = \frac{h c_{0}^{2}}{n^{2}HV}.
\end{equation}
It should be kept in mind that this is the {\em coordinate length} of the wave only.  The physically measurable wavelength is obtained by applying the metric of Eq.~(1).  Thus the measurable wavelength is $\tilde{\lambda} = \lambda / \Phi$.

Using Eq.~(14), we can cast Eq.~(24) into the form
\begin{equation}
\label{lambda}
\lambda = \frac{hc_{0}}{nH\left[1 - m^{2}c_{0}^{4} \Omega^{2}/H^{2}\right]^{1/2}},
\end{equation}
which specifies $\lambda$ entirely in terms of the space functions $n$ and $\Omega$ and the constant of the motion $H$.

\subsection*{B. Index of refraction for massive-particle de Broglie waves}

From Eq.~(\ref{lambda}) we see that the quantity
\begin{equation}
\label{lambdaN}
\lambda n \sqrt{1 - \frac{m^{2}c_{0}^{4} \Omega^{2}}{ H^{2}}} = \frac{hc_{0}}{H}
\end{equation}
is a constant everywhere in the medium.  Now, Eq.~(\ref{lambdaN}), which governs the de 
Broglie waves of the massive particle, is analogous to the familiar rule from ordinary light optics ($\lambda n =$ constant).  Thus the wave-optics rule for massive particles may be expressed
\begin{equation}
\label{lamN}
\lambda N = {\rm constant,} 
\end{equation}
where the index of refraction $N$ for the de Broglie waves of massive particles in the gravitational field is given by
\begin{equation}
\label{N1}
N = n \sqrt{1 - \frac{m^{2}c_{0}^{4} \Omega^{2}}{H^{2}}}.
\end{equation}
Eq.~(\ref{N1}) is an exact result, applying to fields of arbitrary strength. The gravitational wave-optics of massive and massless particles differ only in a dispersion effect, since the index of refraction for massive particles depends on the parameter $m/H$.  From Eq.~(12), note that for light $m/H \rightarrow 0$ and then $N \rightarrow n$. With the use of Eq.~(14), Eq.~(\ref{N1}) may also be written
\begin{equation}
\label{N2}
N = n^{2}V/c_{0},
\end{equation}
as we expected.  With use of Eq.~(16), we also have
\begin{equation}
\label{N3}
N = c_{0}p/H.
\end{equation}
Finally, with use of Eq.~(\ref{hbarw}), the index of refraction may be written explicitly as a function of position and frequency:
\begin{equation}
\label{N5}
N({\mathbf r}, \omega)  = n({\mathbf r}) \sqrt{ 1 - \frac{m^{2}c^{4}_{0}\Omega^{2}({\mathbf r})}{\hbar^{2}\omega^{2}}}.
\end{equation}

By virtue of Eq.~(\ref{N2}), the geodesic equation of motion of the massive particle, 
Eq.~(18), may be written
\begin{equation}
\label{lawofmotion}
\frac{d^{2}{\bf r}}{dA^{2}} = \nabla \left( \frac{1}{2} N^{2}c_{0}^{2}\right), \phantom{XX} {\rm (light\phantom{x}and\phantom{x}particles)}
\end{equation}
and the optical action $A$ may be defined by
\begin{equation}
\label{defA}
\left|\frac{d{\bf r}}{dA}\right| = Nc_{0}. \phantom{XX} {\rm (light\phantom{x}and\phantom{x}particles)}
\end{equation}
These relations are equivalent to Eq.~(18) and Eq.~(19); but now, remarkably, they take on the forms of Eq.~(5) and (6).  All these equations are, it should be stressed, exact 
general-relativistic formulations in the class of metric under consideration.  In the special case of static, isotropic potentials or media, Eq.~(\ref{lawofmotion}) and (\ref{defA}) therefore cover all of the following: classical geometrical optics, classical Newtonian mechanics, general relativistic geometrical optics and general relativistic particle mechanics.  The effective potential energy function is always just the negative square of the index of refraction.

\subsection*{C. Phase and group velocity}

The phase velocity of the waves is $v_{p} = \omega/k = H/p$, or, with use of Eq.~(\ref{N3}), 
\begin{equation}
\label{N4}
v_{p} = c_{0}/N,
\end{equation}
a relation which demonstrates once again that $N$ is playing its proper role as an index of refraction. Inserting $p = \hbar k$ and $H = \hbar \omega$ into the square of Eq.~(\ref{H2}) we obtain
\begin{equation}
\hbar^{2} \omega^{2} = m^{2}c_{0}^{4} \Omega^{2} + c_{0}^{2} \hbar^{2} k^{2}/n^{2}.
\end{equation}
Differentiating with respect to $k$, we obtain
\begin{equation}
\frac{\omega}{k} \frac{d \omega}{d k} = \frac{c_{0}^{2}}{n^{2}},
\end{equation}
or
\begin{equation}
\label{vv1}
v_{p} v_{g} = c_{0}^{2}/n^{2},
\end{equation}
where $v_{g} = d \omega/ d k$ is the group velocity of the de Broglie waves.  Using Eq.~(\ref{N4}), we find 
\begin{equation}
v_{g} = c_{0}N/n^{2}.
\end{equation}
Substituting the expression for $N$ from Eq.~(\ref{N2}), we see that
\begin{equation}
v_{g} = V,
\end{equation}
the coordinate velocity of the classical point-particle.  Using Eq.~(20), we can also write Eq.~(\ref{vv1}) as
\begin{equation}
\label{vv2}
v_{p} v_{g} = c_{0}^{2} \frac{dA}{dt}.
\end{equation}

\subsection*{D. Red shifts of massive particles}
Another way to see that $N$ really is playing the role of an index of refraction for massive particles is to show that it yields the correct gravitational red shifts. Let a particle of mass $m$ travel in free fall between position $\mathbf{r}_1$ and position $\mathbf{r}_2$. The coordinate wavelength $\lambda$ associated with the particle varies according to
\begin{equation}
N(\mathbf{r}_1)\lambda(\mathbf{r}_1) = N(\mathbf{r}_2)\lambda(\mathbf{r}_2).
\end{equation}
The physical or metric length of the waves is $\tilde{\lambda} = \lambda/\Phi$, so the rule governing the redshifts of massive particles in static metrics is
\begin{equation}
N(\mathbf{r}_1)\Phi(\mathbf{r}_1)\tilde{\lambda}(\mathbf{r}_1) = N(\mathbf{r}_2)\Phi(\mathbf{r}_1)\tilde{\lambda}(\mathbf{r}_2),
\end{equation}
or
\begin{equation}
\tilde{\lambda}(\mathbf{r}_1)\sqrt{\Omega^{-2}(\mathbf{r}_1) - c_0^4/H^2} = \tilde{\lambda}(\mathbf{r}_2)\sqrt{\Omega^{-2}(\mathbf{r}_2) - c_0^4/H^2}. 
\end{equation}
Eq.~(45) and (46) are entirely analogous to the rules governing the redshifts of light (for which $n$ stands in place of $N$).\cite{noteredshift} 

\section*{III. WKB SOLUTION TO THE GENERAL \\RELATIVISTIC KLEIN-GORDON EQUATION} 
The index of refraction for massive particles given by Eq.~(\ref{N1}), (\ref{N2}) or (\ref{N3}) has been obtained from semi-classical considerations.  We would like to show that the same expression results from examination of the appropriate quantum-mechanical wave equation.  The Klein-Gordon equation for a spinless massive-particle wavefunction in flat space results from making the substitution $p_{\mu} \rightarrow -i \hbar \partial_{\mu}$ in the Minkowski-space mass-shell constraint $g^{\mu \nu}p_{\mu}p_{\nu} - (mc_{0})^{2} = 0$ and allowing the resulting operators to act on a scalar wavefunction $\psi$.  In the general-relativistic generalization of this wave equation to curved spacetime, the momentum operator must be replaced by the covariant derivative, so that $p_{\mu} \rightarrow -i \hbar \nabla_{\mu}$.  Making this generalization, we arrive at the general relativistic Klein-Gordan equation (GRKGE), a good candidate for a wave equation for massive particles of zero spin:\cite{Bir82}
\begin{equation}
\label{GRKGE}
\left(\, g^{\mu\nu}\,\nabla_{\mu} \nabla_{\nu} + \frac{m^2 \, c_0^2}{\hbar^2}  
\,\right) \psi(x) = 0, 
\end{equation}
where $\psi(x)$ is our scalar field, $g^{\mu\nu}$ is an element of the metric tensor, and we follow the usual convention of implied summation over repeated indices.

Now, the covariant derivative operating on a scalar function just gives the usual partial derivatives:  $\nabla_{\nu} \psi = \partial_{\nu} \psi$.  However, $\partial_{\nu} \psi$ forms a covariant vector.  And when a covariant derivative operator acts on the covariant component $Q_{\nu}(x)$ of a vector, we have
\begin{equation}
\nabla_{\mu} Q_{\nu}(x) = \partial_{\mu} Q_{\nu}(x) - \Gamma^{\lambda}_{\mu \nu}(x)Q_{\lambda}(x),
\end{equation}
where the $\Gamma _{\mu \nu}^{\lambda }$ are the connection coefficients defined by 
\begin{equation}
\Gamma _{\mu \nu}^{\lambda } = \frac{1}{2} g^{\lambda \kappa} \left( \partial_{\mu}g_{\kappa \nu} + \partial_{\nu}g_{\mu \kappa} - \partial_{\kappa}g_{\mu\nu} \right).
\end{equation}

Expanding out the double covariant derivatives, we may write the GRKGE in the more explicit form  
\begin{equation}
\left[\ g^{\mu \nu}\partial _{\mu }\partial _{\nu } -g^{\mu \nu }\Gamma _{\mu \nu}^{\lambda }\partial _{\lambda } + (mc_{0}/\hbar)^{2}\right]\psi(x)~=~0.
\end{equation}
This equation is the jumping-off place for most treatments of scalar matter waves in curved spacetimes.\cite{Don86}

In the static, isotropic metric of Eq.~(1), $g_{\mu \nu} = {\rm diag.} (\Omega^{2}c_{0}^{2}, -\Phi^{-2},-\Phi^{-2}, -\Phi^{-2})$.  Thus the inverse tensor $g^{\mu \nu} = {\rm diag.} (\Omega^{-2}c_{0}^{-2}, -\Phi^{2},-\Phi^{2}, -\Phi^{2})$.  And $g \equiv 
{\rm det}~g_{\mu \nu} = -\Omega^{2}c_{0}^{2}/\Phi^{6}$. 

Thus, with the aid of the identity \cite{Lau65}
\begin{equation}
g^{\mu \nu }\Gamma _{\mu \nu }^{\lambda} =-~(-g)^{-1/2}\partial _{\mu }\left[(-g)^{1/2}g^{\mu \lambda }\right],
\end{equation} 
we obtain  
\begin{equation}
\label{wavepsi}
\frac{n^{2}}{c_{0}^{2}}\frac{\partial ^{2}\psi }{\partial t^{2}}-\mathbf{%
\nabla }^{2}\psi -\mathbf{\nabla }\xi \cdot \mathbf{%
\nabla }\psi +k_{c}^{2} \left( \mathbf{r} \right) \psi =0, 
\end{equation}
where we define
\begin{equation}
\xi(\mathbf{r}) \equiv \ln (\Omega/\Phi )
\end{equation}
and
\begin{equation}
k_{c}\left( \mathbf{r}\right) \equiv mc_{0}/\hbar \Phi.
\end{equation}
$h/mc_{0}$ is the invariant, or physical, Compton wavelength of the particle.  From the considerations in Sec.~II.~D., $\Phi h/mc_{0}$ may be interpreted as a coordinate Compton wavelength. Thus $k_{c}\left( \mathbf{r}\right)$ is the coordinate Compton wave number.

We can remove the term $\mathbf{\nabla }\xi \cdot \mathbf{\nabla }\psi$ from Eq.~(\ref{wavepsi}) by putting 
\begin{equation}
\label{product}
\psi \left( \mathbf{r},t\right) =f(\mathbf{r})\phi \left( \mathbf{r},t\right),   
\end{equation}
where 
\begin{equation}
f(\mathbf{r})= \left(\Phi/\Omega \right)^{1/2} = e^{-\xi/2}.
\end{equation}
Then $\phi \left( \mathbf{r},t\right)$ satisfies the differential equation
\begin{equation}
\label{wave}
\frac{n^{2}}{c_{0}^{2}}\frac{\partial ^{2}\phi }{\partial t^{2}}-\mathbf{%
\nabla }^{2}\phi +\left[ k_{c}^{2}\left( \mathbf{r}\right) +\eta \left( 
\mathbf{r}\right) \right] \phi =0, 
\end{equation}
where we define
\begin{equation}
\label{eta}
\eta \left( \mathbf{r}\right) \equiv \frac{1}{2}\mathbf{\nabla }^{2}\xi
\left( \mathbf{r}\right) +\frac{1}{4}|\mathbf{\nabla }\xi \left( \mathbf{r}%
\right) \mid ^{2}. 
\end{equation}
Note that Eq.~(\ref{wave}) is essentially the flat-spacetime wave equation for a scalar particle in a spatially varying index of refraction $n$ and with a spatially varying Compton wave number $k_{c}$. However, there is an additional term $\eta(\mathbf{r})$ in the wave equation which results from the geometric cross term $g^{\mu \nu }\Gamma _{\mu \nu }^{\lambda}$ in the product of the covariant derivatives. In the weak field limit this term is typically dropped \cite{notedrop} when seeking an approximate solution to the wave equation. However, it is not much extra trouble to carry this term along. In the next section we develop a WKB solution to  Eq.~(\ref{wave}).

\subsection*{A. The WKB expansion}
\label{WKBExpansion}

Let us look for a stationary solution to Eq.~(\ref{wave}) of the form 
\begin{equation}
\label{wavesolution}
\phi \left( \mathbf{r},t\right) =C\exp \left[ \frac{i}{\hbar }\mathcal{S}(\mathbf{r},t)\right]   
\end{equation}
with $C$ constant and where we will take the phase to be of the form 
\begin{equation}
\label{phase}
\mathcal{S}(\mathbf{r},t)=S(\mathbf{r})-H\,t, 
\end{equation}
where $H$ is constant.
Substitution of Eq.~(\ref{wavesolution}) and Eq.~(\ref{phase}) into Eq.~(\ref{wave}) 
shows that we will indeed have a solution, provided that 
\begin{equation}
\label{HJ}
p^{2}-\left( \mathbf{\nabla }S\right) ^{2} +i\hbar \mathbf{\nabla }^{2}S-\hbar ^{2}\,\eta (\mathbf{r})=0,  
\end{equation}
where we have used Eq.~(13) and Eq.~(3).

We now expand $S(\mathbf{r})$ in a power series in $\hbar$: 
\begin{equation}
\label{series}
S(\mathbf{r})=\sum_{n=0}^{\infty }\hbar ^{n}\,S_{n}(\mathbf{r}).  
\end{equation}
Substituting this into Eq.~(\ref{HJ}) yields the set of equations 
\begin{eqnarray}
O(\hbar^{0}):\qquad \left( \mathbf{\nabla }S_{0}\right) ^{2}-p^{2} &=&0
\label{Seq0} \\
O(\hbar^{1}):\qquad 2\mathbf{\nabla }S_{0}\cdot \mathbf{\nabla }S_{1} &=&i%
\mathbf{\nabla }^{2}S_{0}  \label{Seq1} \\
O(\hbar^{2}):\qquad 2\mathbf{\nabla }S_{0}\cdot \mathbf{\nabla }S_{2} &=&i%
\mathbf{\nabla }^{2}S_{1}-\left( \mathbf{\nabla }S_{1}\right)^{2} -\eta %
(\mathbf{r})  \label{Seq2} \\
O(\hbar^{n\geq 3}):\qquad 2\mathbf{\nabla }S_{0}\cdot \mathbf{\nabla }S_{n}&=&i%
\mathbf{\nabla }^{2}S_{n-1}-\sum_{j=1}^{n-1}\mathbf{\nabla }S_{j}\cdot 
\mathbf{\nabla }S_{n-j}.  \label{Seq3}
\end{eqnarray}
 
\subsection*{B. The eikonal equation}
\label{eikonaleqn}
Eq.~(\ref{Seq0}), the eikonal equation, governs the dominant contribution to the phase of the wave.  By direct integration we have
\begin{equation}
\label{S0}
S_{0}(\mathbf{r})=\int^{\mathbf{r}}\mathbf{p(\mathbf{r})}\cdot d\mathbf{x}, 
\end{equation}
where
\begin{equation}
\label{p}
p(\mathbf{r}) \equiv |\mathbf{p}| = \frac{H}{c_0}n\sqrt{1-\left(\frac{\Omega mc_0^2}{H}\right)^2}. 
\end{equation}
This form for $p$ comes from rearrangement of Eq.~(13).

The surface $S_{0}(\mathbf{r})=$ constant is a wavefront with normal in the direction of 
\begin{equation}
\label{fronts}
\mathbf{\nabla }S_{0}(\mathbf{r})=\mathbf{p}(\mathbf{r}). 
\end{equation}
Following Holmes \cite{Hol95}, let us characterize a wavefront by coordinates ${\mathbf{r}} = \bar{\mathbf{x}}(l,\alpha ,\beta )$, where $l$ is the arc length along the ray, normal to surfaces of constant $S_{0}$, and $\alpha$ and $\beta $ are coordinates used to parameterize the wavefront surfaces (e.g., spherical coordinates). $d\overline{\mathbf{x}}/dl$ is the unit vector that is tangent to the ray or trajectory. From Eq.~(16) we see that $d\overline{\mathbf{x}}/dl$ is in the direction of $\mathbf{p}(\mathbf{r})$ and hence, by Eq.~(\ref{fronts}), in the direction of $\mathbf{\nabla }S_{0}(\mathbf{r})$.  Thus  
\begin{equation}
\label{tangent}
\frac{d\overline{\mathbf{x}}}{dl}=\frac{\mathbf{p}(\mathbf{r})}{p}=
\frac{\mathbf{\nabla }S_{0}(\mathbf{r})}{\left| \mathbf{\nabla }S_{0}(\mathbf{r})\right|}. 
\end{equation}
For any $F(\mathbf{r})$ we can write 
\begin{equation}
\label{dF/dl}
\frac{dF}{dl} = \,\frac{d\overline{\mathbf{x}}}{dl}\cdot \mathbf{\nabla }F.
\end{equation}
Putting $S_{0}$ for $F$ in Eq.~(\ref{dF/dl}) and using Eq.~(\ref{tangent}) we have
\begin{equation}
\label{ray}
\frac{dS_{0}}{dl}=\,\frac{d\overline{\mathbf{x}}}{dl}\cdot \mathbf{\nabla }%
S_{0}\,=\, |\mathbf{\nabla }S_{0}|.
\end{equation}
And thus, by Eq.~(\ref{fronts}),
\begin{equation}
\label{S0alt}
S_{0}(l,\alpha ,\beta )=\int^{l}\,p\,dl,  
\end{equation}
which is another way to write Eq.~(\ref{S0}). $\int p\,dl$ is proportional to the optical path length along a segment $dl$ of the ray.  And thus $p$ is proportional to the index of refraction.\cite{Bor80}  So here we have a more detailed justification of 
Eq.~(\ref{N3}). 

\subsection*{C. The transport equation}
\label{transporteqn}

We now need to solve the transport equation, Eq.~(\ref{Seq1}). Putting $S_{1}$ for $F$ in Eq.~(\ref{dF/dl}), we have
\begin{equation}
p\frac{dS_{1}}{dl} = \mathbf{\nabla }S_{0} \cdot \mathbf{\nabla }S_{1},
\end{equation}
where we have used Eq.~(\ref{fronts}).  Then, by substitution, Eq.~(\ref{Seq1}) becomes
\begin{equation}
\frac{dS_{1}}{dl} = \frac{i}{2p}\nabla^{2}S_{0},
\end{equation}
with solution 
\begin{equation}
\label{S1}
S_{1}(l,\alpha ,\beta )=\frac{i}{2}\int^{l}dl\frac{\mathbf{\nabla }^{2}S_{0}}{p}. 
\end{equation}
A short calculation shown in the Appendix reveals that
\begin{equation}
\label{theorem}
\frac{\mathbf{\nabla }^{2}S_{0}}{p}=\frac{d(p\,J)/dl}{p\,J}, 
\end{equation}
where 
\begin{equation}
\label{J}
J=\left| \frac{\partial \mathbf{x}}{\partial (l,\alpha ,\beta )}\right|
\end{equation}
is the Jacobian of the transformation from the curvilinear ray coordinates 
$(l,\alpha ,\beta )$ to Cartesian coordinates. Then, upon substitution of Eq.~(\ref{theorem}) into Eq.~(\ref{S1}), one can perform the integration to obtain 
\begin{equation}
\label{eq44}
S_{1}\left(\mathbf{r(}l,\alpha ,\beta )\right)=\frac{i}{2}\ln \left( \frac{p(\mathbf{r}%
)J(\mathbf{r})}{p(\mathbf{r}_{0})J(\mathbf{r}_{0})}\right) = \frac{i}{2}\mu (\mathbf{r}),  
\end{equation}
where $\mu (\mathbf{r})\equiv \ln \left( \frac{p(\mathbf{r})J(%
\mathbf{r})}{p(\mathbf{r}_{0})J(\mathbf{r}_{\mathbf{0}})}\right)$ and 
$\mathbf{r}_{\mathbf{0}} \equiv \mathbf{r}|_{l=0}$. To the lowest order ($O(\hbar^0)$ in the phase and amplitude), we have found the WKB approximate solution 
\begin{equation}
\label{eq45} 
\psi _{1}\left( \mathbf{r},t\right) = C\sqrt{\frac{\Phi (\mathbf{%
r})}{\Omega (\mathbf{r})}}\sqrt{\frac{p(\mathbf{r}_{0})J(\mathbf{r}_{0})}{p(%
\mathbf{r})J(\mathbf{r})}}\exp \left[ \frac{i}{\hbar}\left( \int^{%
\mathbf{r}}\mathbf{p(\mathbf{r})}\cdot d\mathbf{x}-H\,t\right) \right] ,  
\end{equation}
where the subscript $1$ on $\psi _{1}\left( \mathbf{r},t\right) $ indicates
that we have carried out the WKB expansion to $S_{n=1}.$ 

Note that Eq.~(\ref{eq45}) takes into account, to lowest order, the term 
$\xi \mathbf{(r)}\equiv \ln (\Omega/\Phi)$, which arises from the covariant derivative term $g^{\mu \nu }\Gamma _{\mu \nu }^{\lambda }$. What has been left out are higher order terms involving $\mathbf{\nabla }^{2}\xi \left( \mathbf{r}\right)$ and 
$\mathbf{\nabla }\xi \left( \mathbf{r}\right) $ in $\eta \left( \mathbf{r}\right) $, which come from the quantum corrections terms, $S_{n\geq 2}$ (which we deal with next). Note also that this solution is valid for strong gravitational fields.\cite{Don86} But truncating the solution at $S_{1}$ does impose the restriction that $p^{2}\gg \hbar^{2} \eta\left(\mathbf{r}\right)$, or
\begin{equation}
\label{eq46}
\lambda |\mathbf{\nabla }\xi| ,\ \lambda|\sqrt{\mathbf{\nabla}^{2}\xi 
}| \ll 1.
\end{equation}
Since $\xi \mathbf{(r)}$ is slowly varying, the above conditions can hold even reasonably close to the horizon of a black hole. We investigate this in the next section.

\subsection*{D. Estimation of terms in Schwarzschild metric}

If we had defined $\phi \left( \mathbf{r},t\right) =u\left( \mathbf{r}%
\right) \exp( -i H t/\hbar) $ our wave equation, Eq.~(\ref{wave}), would have the form of the Helmholtz equation: 
\begin{equation}
\label{eq47}
\mathbf{\nabla }^{2}u(\mathbf{r})+k^{2}(\mathbf{r})\left[ 1-\frac{\eta (%
\mathbf{r})}{k^{2}(\mathbf{r})}\right] u(\mathbf{r})=0 
\end{equation}
where $p(\mathbf{r})=\hbar \,k(\mathbf{r}),\,k(\mathbf{r})=k_{0}\,N,$ with 
$k_{0}=\omega _{0}/c_{0}=H/\hbar c_{0}.$ Therefore we want to consider the
order of magnitude of the terms $|\mathbf{\nabla }\xi|^{2}/k^{2}$ and 
$|\mathbf{\nabla }^{2}\xi|/k^{2}$. 

For the Schwarzschild metric in isotropic coordinates we have:
\begin{eqnarray}
\Omega  &=&\frac{1-1/\rho }{1+1/\rho } \nonumber \\
\Phi &=&\frac{1}{\left( 1+1/\rho\right) ^{2}} \nonumber \\
n &=&\frac{\left( 1+1/\rho \right) ^{3}}{\left( 1-1/\rho \right)}\nonumber \\
\xi &=&\ln (1- 1/\rho ^{2}) \nonumber \\
k&=&k_{0}n\sqrt{1- \Omega ^{2}/H^{\prime 2}},  \label{Sch}
\end{eqnarray}
where $H^{\prime 2}\equiv H/(mc_{0}^{2})$ and $\rho =r/r_{s},$ with 
$r_{s}= GM/c_{0}^{2}=0.74\,M/M_{\odot }$ km (half 
the Schwarzschild radius). 

For massless particles, $H^{\prime }\rightarrow \infty $, so that $k=k_{0}n$ and
we obtain 
\begin{eqnarray}
\frac{|\mathbf{\nabla }\xi \mathbf{(r)}|^{2}}{k^{2}(\mathbf{r
})}=\frac{4}{\left( k_{0}r_{s}\right) ^{2}}\frac{1}{\rho ^{6}}\frac{1}{%
\left( 1+1/\rho \right) ^{8}} \nonumber \\
\frac{|\mathbf{\nabla }^{2}\xi \mathbf{(r)}|}{k^{2}(\mathbf{r})}=\frac{2}{\left( k_{0}r_{s}\right) ^{2}}\frac{1}{\rho^{4}}\frac{\left( 1+1/\rho ^{2}\right) }{\left( 1+1/\rho \right) ^{8}}.
\label{Sch2}
\end{eqnarray}
These ratios are finite for all values $1\leq \rho <\infty $ (the valid range of the isotropic scaled radius $\rho$), even though the numerators and denominators of the left hand sides of Eq.~(\ref{Sch2}) each separately diverge as $\rho\rightarrow 1$. From Eq.~(\ref{Sch}), $k\mathbf{(\rho)} \rightarrow   k_{0}$ as $\rho\rightarrow \infty$, so $k_{0}$ is the usual wave number of the particle at spatial infinity. For ordinary wavelengths, 
$k_{0}r_{s} \gg 1$, since $r_{s}\sim $ km, so the above ratios remain very small even down to $\rho=1$. Even if we were to consider ultra-long wavelengths such that $k_{0}r_{s}\sim 1,$ the ratios in Eq.~(\ref{Sch2}) could still be made small for values of $\rho \sim 2$, i.e. 
$r=2r_{s}$, the Schwarzschild radius. In this case, the term $\eta (\mathbf{r})$ arising from the covariant derivative term $g^{\mu \nu }\Gamma _{\mu \nu }^{\lambda }$ is negligible compared to $k^2(\mathbf{r})$ all the way down to the Schwarzschild radius, and for all intents and  purposes, our wave equation for massless particles is of the form 
\begin{equation}
\label{eq50}
\mathbf{\nabla }^{2}u(\mathbf{r})+k^{2}(\mathbf{r})\,u(\mathbf{r})=0.
\end{equation}
However, in the next section it causes no great difficulty to carry terms involving $\eta(\mathbf{r})$ along formally.

\subsection*{E. Quantum corrections}
\label{qcorrections}

The first occurrence of $\eta \left( \mathbf{r}\right)$ appears in Eq.~(\ref{Seq2}). Using Eq.~(\ref{dF/dl}) we can write a first order equation for $dS_{2}/dl$ whose solution is given by 
\begin{equation}
\label{S2}
S_{2}(l,\alpha ,\beta )=-\int^{l}\frac{dl}{2p}\left[ \frac{1}{2}\mathbf{%
\nabla }^{2}\mu \mathbf{(r)-}\frac{1}{4}\left( \mathbf{\nabla }\mu \mathbf{%
(r)}\right) ^{2}+\eta \left( \mathbf{r}\right) \right]. 
\end{equation}
Since $S_{2}$ is purely real, this is an $O(\hbar ^{2}\mathbf{)}$ correction to the phase.

The remaining equations, Eq.~(\ref{Seq3}), can be formally solved to give 
\begin{equation}
\label{Sn}
S_{n}(l,\alpha ,\beta )=\int^{l}\frac{dl}{2p}\left[ i\mathbf{\nabla }%
^{2}S_{n-1}-\sum_{j=1}^{n-1}\mathbf{\nabla }S_{j}\cdot \mathbf{\nabla }%
S_{n-j}\right], 
\end{equation}
where the terms in the brackets involve only the previously determined functions 
$ S_{1},S_{2},\ldots ,S_{n-1}$. The first correction to the amplitude is $O(\hbar ^{2}\mathbf{)}$ and is given by 
\begin{equation}
\label{eq53}
S_{3}(l,\alpha ,\beta )=i\int^{l}\frac{dl}{2p}\left[ \mathbf{\nabla }%
^{2}S_{2}\mathbf{-\mathbf{\nabla }}\mu \mathbf{\mathbf{(r)\cdot }\nabla }%
S_{2}\right] \equiv i S_{3}'.  
\end{equation}
Putting this all together, we have to $O(\hbar ^{2}\mathbf{)}$ in the phase and amplitude 
\begin{eqnarray}
\label{eq54}
\psi _{3}\left( \mathbf{r},t\right) &=&C\sqrt{\frac{\Phi (\mathbf{r})}{\Omega (\mathbf{r})}}\sqrt{\frac{p(\mathbf{r}_{0})J(\mathbf{r}%
_{0})}{p(\mathbf{r})J(\mathbf{r})}}\exp \left( -\hbar ^{2}S_{3}'(\mathbf{r})\right) \times
 \nonumber \\
&&\exp \left[ \frac{i}{\hbar }\left( \int^{\mathbf{r}}\mathbf{p(r)}\cdot d%
\mathbf{x-}H\,t + \hbar^{2}S_{2}(\mathbf{r})\right) \right], 
\end{eqnarray}
where the subscript $3$ on $\psi _{3}\left( \mathbf{r},t\right) $ indicates
that we have carried out the WKB expansion to $S_{n=3}.$

The lowest-order contribution to the phase of the wavefunction in Eq.~(\ref{eq54}) is just
$-\hbar^{-1}$ times the classical action 
\begin{equation}
\label{eq55}
S_{cl} = \int\,p_{\mu}\,dx^{\mu}, 
\end{equation} 
where $p_{\mu} = (H/c_0,\mathbf{p})$.  If this four-vector is substituted into the mass shell constraint $g^{\mu\nu}p_{\mu}p_{\nu}-(m c_0)^2=0$, we obtain the square of the Hamiltonian, Eq.~(\ref{H2}). Using  Eq.~(\ref{N3}) and the definition of $V=dl/dt$, we can also write the classical action as $S_{cl}(\mathbf{r})= H/c_0\,\int^{\mathbf{r}} \left(  N - c_0/V \right) dl$ along the geodesic.  For massless particles, $N\to n$ and $V\to c_0/n$ so that $S_{cl}\to 0$. Thus, the quantum correction term $S_2(\mathbf{r})$ in  Eq.~(\ref{eq54}) become the lowest order contribution to the phase of the wavefunction in this case.

In 1979, Stodolsky \cite{Gre79} argued that the wave function of a spinless quantum mechanical particle in a arbitrary metric $ds^2=g_{\mu\nu}dx^{\mu}dx^{\nu}$ should take the form $\psi(\mathbf{r},t) = A\,\exp[-iS_{cl}(\mathbf{r},t)/\hbar]$, with $S_{cl}$ given by Eq.~(\ref{eq55}).  This expression for the phase of a quantum mechanical wavefunction in curved spacetime have been used recently to compute the effects of mass oscillations from supernova neutrinos \cite{neutrinos}. Eq.~(\ref{eq54}) reveals that  the phase of the wavefunction for a massive spin $0$ particle only has the form given by Eq.~(\ref{eq55}) to lowest order in $\hbar$. The Klein-Gordon equation, which is second order in space, cannot have a solution in which the phase is given exactly by Eq.~(\ref{eq55}), since two spatial derivatives of such a phase would produce terms like $\mathbf{\nabla}\cdot\mathbf{p}$, which are not present in the wave equation. It is interesting to note that the wave function for a spin $1/2$ particle in a static metric has \textit{precisely} the form proposed by Stodolsky, Eq.~(\ref{eq55}) \cite{QMphaseinGR}. This result follows from the first-order nature of the Dirac equation in curved spacetime.

\section*{IV. THE OPTICAL ACTION}
\subsection*{A. The optical action and the proper time}

Eq.~(\ref{defA}) defines the optical action for both massive and massless particles.  Eq.~(7), which also is valid for both massive and massless particles, provides the relation between the optical action and the coordinate time elapsed along a trajectory.  It will be convenient to state explicitly the connection between $A$ and the most commonly used parameter for massive particle trajectories, the proper time $\tau$.  From Eq.~(\ref{metric}) and Eq.~(12), the relation between the elapsed proper time $d\tau (= ds/c_{0})$ and the optical action $dA$ may be written
\begin{equation}
\label{Atau}
dA = \frac{\Phi^{2}H}{mc_{0}^{2}}\,d\tau.
\end{equation}
$\tau$ is not a suitable parameter for light rays since, for massless particles, $d\tau \rightarrow 0$.  However, in this case $H/m \rightarrow \infty$ in such a way that the product $(H/m) d\tau$ remains finite, as we see from Eq.~(7).  This is why the o
ptical action may be used for parameterizing the trajectories of both massive and massless particles.

\subsection*{B. The optical action in the WKB calculation}
\label{OpticalAction}
The optical action $dA$ is related very simply to the measure $dl/p$, which appears in
the integrals for the higher order WKB phase terms in Eq.~(\ref{S1}), 
Eq.~(\ref{S2}) and Eq.~(\ref{Sn}). From the definition of $A$ [Eq.~(\ref{defA})], we have $dl = Nc_{0}dA$.  Using Eq.~(\ref{N3}), we obtain 
\begin{equation}
\label{intmeas}
dA=\frac{H}{c_{0}^{2}p}dl. 
\end{equation}
Thus, the phase in the expressions for $S_{n\geq 1}$ is integrated with respect to the optical action. 

\subsection*{C. The optical action as Born and Wolf's $\tau'$}
In their discussion of geometrical optics, Born and Wolf\cite{tau} introduce an operator 
\begin{equation}
\label{op}
\frac{\partial }{\partial \tau'}=\mathbf{\nabla}S_0\cdot\mathbf{\nabla},
\end{equation}
where $\tau'$ is a certain parameter that specifies position along a light ray. (Born and Wolf actually write $\tau$ rather than $\tau'$.  We add the $'$ to avoid confusing this parameter with the proper time in general relativity.) Applying this operator to the eikonal $S_0$ itself gives 
\begin{equation}
\label{eq72}
\frac{dS_0}{d\tau'}=\left( \mathbf{\nabla}S_0\right)
^{2}=p^{2}, 
\end{equation}
and thus
\begin{equation}
dS_0=p^{2}\,d\tau'.  
\end{equation}
However, from Eq.~(\ref{S0alt}) we have 
\begin{equation}
dS_0=p\,dl.  
\end{equation}
Comparing these two expressions, we see that 
\begin{equation}
\label{eq74}
d\tau' =\frac{dl}{p}. 
\end{equation}
From this it follows that the operator defined by Eq.~(\ref{op}) is $p \,\partial_{l}$, that is, $p$ times the derivative along the ray. Comparing Eq.~(\ref{eq74}) with Eq.~(\ref{intmeas}), we see that, apart from a factor of $H/c_{0}^{2}$, Born and Wolf's $d \tau'$ is the optical action $dA$. 
 
\subsection*{D. The optical action as a wave-and-particle action} Eq.~(\ref{vv2}) also provides some insight into the significance of the optical action.  Multiplying Eq.~(\ref{vv2}) by $dt$ we have
\begin{equation}
dA = v_{p} \, (v_{g} dt)/ c_{0}^{2} = v_{p} \, dl/ c_{0}^{2},
\end{equation}
or in words (ignoring the factor of $c_{0}^{2}$),
\begin{quote}
change in {\it A} = phase speed of waves $\times$ displacement of particle.
\end{quote}
Alternatively, this may be written 
\begin{equation}
dA = v_{g} \, (v_{p} dt) / c_{0}^{2},
\end{equation}
\begin{quote}
change in {\it A} = speed of particle $\times$ displacement of phase wave.
\end{quote}
These expressions for $dA$ bear a rather intriguing relation to the classical 
action of a Newtonian particle ($=$ speed of particle $\times$ displacement of 
particle).  The optical action $A$ for a massive particle has, as it were, a 
foot in the wave regime and a foot in the particle regime. 

\section*{V. SUMMARY}
In a static, isotropic metric, it is possible to define an index of refraction not only for massless, but also for massive particles.  The index of refraction $N$ is given by Eq.~(\ref{N1}), (\ref{N2}) or (\ref{N3}).  This index of refraction works as it should for de Broglie waves in the physical-optics rules of Eq.~(\ref{lamN}) and Eq.~(\ref{N4}). Thus one may work out physical-optics applications, including interference and diffraction, using this $N$.\cite{Gre79} Passing over from physical optics to geometrical optics, we attain the point-particle regime.  In the point-particle regime, the trajectory of the particle (whether massive or massless) is expressed by an equation of motion formally identical to Newton's second law, Eq.~(\ref{lawofmotion}), in which the independent variable is the optical action, defined by Eq.~(\ref{defA}), and in which the potential energy function is $-N^{2}c_{0}^{2}/2$.  We have confirmed this index of refraction, and explored the limits of its validity, by giving an explicit WKB solution to the general-relativistic Klein-Gordon equation in three spatial dimensions.  

\section*{APPENDIX: PROOF OF EQ.~(\ref{theorem})}
\label{appendixB}
In this appendix we will prove Eq.~(\ref{theorem}):
\begin{displaymath}
\label{B0} 
\frac{\mathbf{\nabla }^{2}S_{0}}{p}=\frac{d\ln \left( pJ\right)}{dl} 
\end{displaymath}
where $J$ is the Jacobian of the transformation from the curvilinear coordinates 
$l,\alpha,\beta$ (which are used to describe the wave fronts $S_0(\mathbf{r}) =$ constant) to Cartesian coordinates. In order prove this relation, we must first prove the following lemma:
\begin{equation}
\frac{dJ}{dl}= J\,\mathbf{\nabla \cdot }\left( \frac{\mathbf{\nabla }S_{0}}{p}\right) =J\,\mathbf{\nabla \cdot }\frac{d\overline{\mathbf{x}}}{dl},
\label{B1}
\end{equation}
which states that the logarithmic derivative of the Jacobian $J$ along a congruence of  trajectories is equal to the divergence of the tangent vector field of the congruence. 

A standard result, proven in many relativity texts\cite{DIn92}, is that for any matrix 
$a_{ij}$, with determinant $a$ and inverse $a^{ij}=A^{ji}/a$, where $A^{ij}$ is the signed cofactor of $a_{ij}$, we have 
\begin{equation}
 \label{B2}
\frac{\partial a}{\partial x^{k}}=a\,a^{ji\,}\frac{\partial a_{ij}}{\partial
x^{k}}= a\,a^{ij\,}\frac{\partial a_{ij}}{\partial x^{k}}%
\,\,\,\,\textrm{for }a_{ij}\textrm{ symmetric.} 
\end{equation}
Let 
\begin{equation}
J=\left| \frac{\partial \mathbf{x}}{\partial (l,\alpha ,\beta )}\right|
\label{B3}
\end{equation}
be the determinant of the transformation matrix from the curvilinear ray coordinates 
$(l,\alpha ,\beta )$ to the Cartesian coordinates $\mathbf{x}$. Writing out Eq.~(\ref{B2}) with $a_{ij}\rightarrow J_{ij}=\partial x^{i}/\partial x^{\prime j}$, with $x^{i}$ as Cartesian coordinates and $x^{\prime i}$ as curvilinear coordinates yields 
\begin{equation}
\frac{\partial J}{\partial x^{^{\prime }k}}=J\,\frac{\partial x^{\prime j}}{%
\partial x^{i}}\frac{\partial }{\partial x^{\prime k}}\left( \frac{\partial
x^{i}}{\partial x^{^{\prime }j}}\right) =J\,\frac{\partial x^{\prime j}}{%
\partial x^{i}}\frac{\partial }{\partial x^{\prime j}}\left( \frac{\partial
x^{i}}{\partial x^{^{\prime }k}}\right) =J\,\frac{\partial }{\partial x^{i}}%
\left( \frac{\partial x^{i}}{\partial x^{^{\prime }k}}\right) ,  \label{B4}
\end{equation}
where in the second step we have switched the order of partial differentiation, and in the third step we have used the chain rule. If we set $k=1$ with $x^{^{\prime }1}=l$, the expression in
the last parentheses above is $d\mathbf{x}/dl$, in component form. Thus, with $k=1$, Eq.~(\ref{B4}) is just Eq.~(\ref{B1}) in component form.

Now, using Eq.~(\ref{B1}), we want to prove Eq.~(\ref{theorem}).
We begin with the first equality in Eq.~(\ref{B1}) and expand out the 
divergence to obtain 
\begin{equation}
\frac{d\ln J}{dl}=-\frac{\mathbf{\nabla }S_{0}}{p}\cdot \frac{\mathbf{\nabla 
}p}{p}+\frac{\mathbf{\nabla }^{2}S_{0}}{p}. 
\end{equation}
Using the ray equation, Eq.~(\ref{tangent}), in the form 
\begin{equation}
\frac{d\overline{\mathbf{x}}}{dl}=\frac{\mathbf{\nabla }S_{0}(\mathbf{r})}{p},
\label{B6}
\end{equation}
we get
\begin{equation}
\frac{d\ln J}{dl}=-\frac{d\overline{\mathbf{x}}}{dl}%
\cdot \mathbf{\nabla }\ln p+\frac{\mathbf{\nabla }^{2}S_{0}}{p},
\end{equation}
or, with use of Eq.~(\ref{dF/dl}),
\begin{equation}
\label{B7}
\frac{d\ln J}{dl}= -\frac{d\ln p}{dl}+\frac{\mathbf{\nabla }^{2}S_{0}}{p}.
\end{equation}
Solving for $\mathbf{\nabla }^{2}S_{0}/p$ in Eq.~(\ref{B7}) yields 
the desired result, Eq.~(\ref{theorem}).

\newpage

\expandafter\ifx\csname bibnamefont\endcsname\relax
  \def\bibnamefont#1{#1}\fi
\expandafter\ifx\csname bibfnamefont\endcsname\relax
  \def\bibfnamefont#1{#1}\fi
\expandafter\ifx\csname url\endcsname\relax
  \def\url#1{\texttt{#1}}\fi
\expandafter\ifx\csname urlprefix\endcsname\relax\def\urlprefix{URL }\fi
\providecommand{\bibinfo}[2]{#2}
\providecommand{\eprint}[2][]{\url{#2}}


\begin{thebibliography}{99}

\bibitem{noteiso}
\bibinfo{note}{For a systematic technique for finding isotropic coordinates, see R.~Adler, M.~Bazin, and M.~Schiffer, {\it Introduction to General Relativity and Gravitation} (McGraw-Hill, New York, 1965), pp.~174-177. Many examples of metrics that can be
 put into isotropic form are given in Ref.\cite{noteENI1}.}

\bibitem{noteEddington}
\bibinfo{note}{The idea of representing the gravitational field as an optical medium with an effective index of refraction goes all the way back to the early days of general relativity.  A.~S.~Eddington, {\it Space, Time and Gravitation} (Cambridge U.\ P.
, Cambridge, 1920), p.~109.}

\bibitem{noteER}
\bibinfo{note}{This optical theory has a long history, going back to Bernoulli and Maxwell.  For a systematic development see J.~Evans and M.~Rosenquist, ```$F = ma$' {\it optics},'' Am.\ J.\ Phys.\ {\bf 54}, 876-883 (1986) and J.~Evans, ``Simple forms for equations of rays in 
gradient-index lenses,'' Am.\ J.\ Phys.\ {\bf 58}, 773-778 (1990).} 

\bibitem{Nan95}
  \bibinfo{author}{\bibfnamefont{K.~K.}~\bibnamefont{Nandi}} \bibnamefont{and}
  \bibinfo{author}{\bibfnamefont{A.} \bibnamefont{Islam}},
  \bibinfo{title}{``On the optical-mechanical analogy in general relativity,''}
  \bibinfo{journal}{Am. J. Phys.} \textbf{\bibinfo{volume}{63}},
  \bibinfo{pages}{251-256} (\bibinfo{year}{1995}).

\bibitem{noteENI1}
  \bibinfo{note}{This section briefly summarizes results given in J.~Evans, K.~K.~Nandi and A.~Islam, ``On the optical-mechanical analogy in general relativity: Exact Newtonian forms for the equations of motion of particles and photons,'' Gen.\ Rel.\ Grav
.\ {\bf 28}, 413-439 (1996).}

\bibitem{noteENI2}
\bibinfo{note}
{For a simpler introduction, with examples based on the Schwarzschild metric, see J.~Evans, K.~K.~Nandi and A.~Islam, ``The optical-mechanical analogy in general relativity: New methods for the paths of light and of the planets,'' Am.\ J.\ Phys.\ {\bf 64}
, 1404-1415 (1996).}

\bibitem{noteAlsing}
\bibinfo{note}{For an extension of the theory to a broader class of metrics see P.~M.~Alsing, ``The optical-mechanical analogy for stationary metrics in general relativity,'' Am.\ J.\ Phys.\ {\bf 66}, 779-790 (1998).}

\bibitem{noteBro}
  \bibinfo{note}{Louis de Broglie, {\it Recherches sur la th\'{e}orie des quanta} (Masson, Paris, 1963), a reprint of the thesis of 1924.} 

\bibitem{noteredshift}
  \bibinfo{note}{See Ref.\cite{noteENI1}, pp.~435-437.}

\bibitem{Bir82}
\bibinfo{note}{For problems concerning quantization in curved space see the
  discussions in}
\bibinfo{author}{\bibfnamefont{N.}~\bibnamefont{Birrell}} \bibnamefont{and}
  \bibinfo{author}{\bibfnamefont{P.}~\bibnamefont{Davies}},
  \emph{\bibinfo{title}{Quantum Fields in Curved Space}}
  (\bibinfo{publisher}{Cambridge U. P.}, \bibinfo{address}{New York},
  \bibinfo{year}{1982})
\bibinfo{note}{and in S.A. Fulling, ``Nonuniqueness of
  canonical field quantization in a Riemannian space-time,'' Phys. Rev. D.
  \textbf{7}, 2850-2862 (1973); D.W. Sciama, P. Candellas and D. Deutsch,
  ``Quantum field theory, horizons and thermodynamics,'' Advances in
  Physics \textbf{30}, 327-366 (1981); S. A. Fulling, \textit{Aspects of
  Quantum Field Theory in Curved Space-Time}, (Cambridge U. P., New York,
  1989).}

\bibitem{Don86}
\bibinfo{note}{A solution for the wavefunction in the limit of weak gravitational fields is given by}
\bibinfo{author}{\bibfnamefont{J.}~\bibnamefont{Donoghue}} \bibnamefont{and}
  \bibinfo{author}{\bibfnamefont{B.}~\bibnamefont{Holstein}},
  \bibinfo{title}{``Quantum mechanics in curved space,''}
  \bibinfo{journal}{Am. J. Phys.} \textbf{\bibinfo{volume}{54}},
  \bibinfo{pages}{827-831} (\bibinfo{year}{1986}).

\bibitem{Lau65}
\bibinfo{author}{\bibfnamefont{L.}~\bibnamefont{Landau}} \bibnamefont{and}
  \bibinfo{author}{\bibfnamefont{E.}~\bibnamefont{Lifshitz}},
  \emph{\bibinfo{title}{The Classical Theory of Fields}}
  (\bibinfo{publisher}{Addison-Wesley}, \bibinfo{address}{Reading,
  Massachusetts}, \bibinfo{year}{1965}),
  \bibinfo{pages}{p.~244}.

\bibitem{notedrop}
  \bibinfo{note}{For example, in Ref.~\cite{Don86}.}

\bibitem{Hol95}
\bibinfo{author}{\bibfnamefont{M.~H.} \bibnamefont{Holmes}},
  \emph{\bibinfo{title}{Introduction to Perturbation Methods}}
  (\bibinfo{publisher}{Springer-Verlag}, \bibinfo{address}{New York},
  \bibinfo{year}{1995}),
  \bibinfo{pages}{pp.~197-207}.

\bibitem{Bor80}
\bibinfo{author}{\bibfnamefont{M.}~\bibnamefont{Born}} \bibnamefont{and}
  \bibinfo{author}{\bibfnamefont{E.}~\bibnamefont{Wolf}},
  \emph{\bibinfo{title}{Principles of Optics,}} 
  (\bibinfo{publisher}{Pergamon Press}, \bibinfo{address}{New York},
  \bibinfo{year}{1990}),
   \bibinfo{pages}{6th ed., pp.~114-115.}

\bibitem{Gre79}
  \bibinfo{note}{For an introduction to this subject, see}
  \bibinfo{author}{\bibfnamefont{D.}~\bibnamefont{Greenberger}} \bibnamefont{and}
  \bibinfo{author}{\bibfnamefont{A.~W.} \bibnamefont{Overhauser}},
  \bibinfo{title}{``Coherent effects in neutron diffraction and gravity experiments,''}
  \bibinfo{journal}{Rev. Mod. Phys.} \textbf{\bibinfo{volume}{51}},
  \bibinfo{pages}{43-78} (\bibinfo{year}{1979})
  \bibinfo{note}{and}
\bibinfo{author}{\bibfnamefont{L.}~\bibnamefont{Stodolsky}},
  \bibinfo{title}{``Matter and light wave interferometry in gravitational 
fields,''}
  \bibinfo{journal}{Gen. Rel. Grav.} \textbf{\bibinfo{volume}{11}},
  \bibinfo{pages}{391-405} (\bibinfo{year}{1979}).

\bibitem{neutrinos}
\bibinfo{author}{\bibfnamefont{N.}~\bibnamefont{Fornengo}},
  \bibinfo{author}{\bibfnamefont{C.}~\bibnamefont{Giunti}},
  \bibinfo{author}{\bibfnamefont{C.~W.} \bibnamefont{Kim}}, \bibnamefont{and}
  \bibinfo{author}{\bibfnamefont{J.}~\bibnamefont{Song}},
  \bibinfo{title}{``Graviational effects on the neutrino osciallations,''}
  \bibinfo{journal}{Phys. Rev. D} \textbf{\bibinfo{volume}{56}},
  \bibinfo{pages}{1895-1902} (\bibinfo{year}{1997});
 \bibinfo{author}{\bibfnamefont{T.}~\bibnamefont{Bhattacharya}},
  \bibinfo{author}{\bibfnamefont{S.}~\bibnamefont{Habib}}, \bibnamefont{and}
  \bibinfo{author}{\bibfnamefont{E.}~\bibnamefont{Mottola}},
  \bibinfo{title}{``Gravitationally Induced Neutrino-Oscillation Phases in Static Spacetimes,''}
  \bibinfo{journal}{Phys. Rev. D} \textbf{\bibinfo{volume}{59}},
  \bibinfo{pages}{067301, 4 pages} (\bibinfo{year}{1999}).

\bibitem{QMphaseinGR}
\bibinfo{author}{\bibfnamefont{P.~M.}~\bibnamefont{Alsing}},
  \bibinfo{author}{\bibfnamefont{J.~C.}~\bibnamefont{Evans}}, \bibnamefont{and}
  \bibinfo{author}{\bibfnamefont{K.~K.}~\bibnamefont{Nandi}},
  \bibinfo{title}{``The phase of a quantum mechanical particle in curved spacetime,''}
  \bibinfo{journal}{Gen. Rel. Grav.} (\bibinfo{year}{in press for Sept., 2001}).
\bibinfo{note}{Preprint at http://arXiv.org,  gr-qc0010065 (2000).}

\bibitem{tau}
\bibinfo{note}{Ref.\cite{Bor80}, Eq.~(38), p.~117.}

\bibitem{DIn92}
\bibinfo{author}{\bibfnamefont{R.}~\bibnamefont{D'Inverno}},
  \emph{\bibinfo{title}{Introducing Einstein's Relativity}}
  (\bibinfo{publisher}{Clarendon Press}, \bibinfo{address}{New York},
  \bibinfo{year}{1992}),
  \bibinfo{pages}{pp.~93-94}.  
\bibinfo{note}{See also Ref.\cite{Lau65}, pp.~242-243.}


\end{thebibliography}
\end{document}